This pre-print has been accepted by the 2026 IEEE International Conference on Intelligent Transportation Systems (ITSC). Naples, Italy.

# Agentic Vehicles for Human-Centered Mobility: Definition, Prospects, and Synergistic Co-Development with Vehicle Autonomy

Jiangbo Yu*, Raphaël Frank*, Luis Miranda-Moreno, Sasan Jafarnejad, Jônatas Augusto Manzolli, Fuqiang Liu, Jiyao Wang, Pavel Chernakov, Ali Eslami

***Abstract*—Autonomy, from the Greek *autos* (self) and *nomos* (law), refers to the capacity to operate according to internal rules without external control. Autonomous vehicles (AuVs) are therefore understood as vehicular systems that perceive their environment and execute tasks with minimal human intervention, consistent with the direction indicated by the SAE levels of automated driving. However, recent research and deployments increasingly showcase vehicular capabilities that, while not contradicting autonomy, are not entailed by it, including ambiguous goal handling, purposeful social engagement, external tool use, proactive problem solving, continuous learning, and context-sensitive reasoning in unseen and ethically salient situations, enabled in part by multimodal language models. These developments reveal a gap between technical autonomy and the broader social cognitive functions required for human-centered mobility, which are more precisely captured by the notion of *agency*. Therefore, rather than adding increasingly elaborate modifiers to "autonomous," we introduce agentic vehicles (AgVs) and suggest that autonomy and agency are intertwined but conceptually distinct: if autonomy concerns what to do and how to do it (task executions under internal rules), agency pertains to why to do it and what else can be done (goal-directed, adaptive actions). We further present autonomy and agency as orthogonal yet synergistic dimensions with co-development implications. Vehicle agency marks a novel dimension of mobility service intelligence, heralding vehicles as purposeful actors in society.**

*The authors based in Canada appreciate the financial support from the Fonds de recherche du Québec (FRQ) AUDACE program, the New Frontiers in Research Fund (NFRF) administered by the Canada Research Coordination Committee (CRCC), and the Natural Sciences and Engineering Research Council of Canada (NSERC) Discovery Grant. This work was also supported in part by the Luxembourg National Research Fund (FNR), grant reference: INR/AUDACE/25/19343746/EAMCIMCA.*

**Co-corresponding authors: Jiangbo Yu (jiangbo.yu@mcgill.ca) and Raphaël Frank (raphael.frank@uni.lu).*

*Raphaël Frank, Sasan Jafarnejad, Pavel Chernakov are with the Interdisciplinary Centre for Security, Reliability and Trust (SnT), University of Luxembourg, L-4365 Esch-sur-Alzette, Luxembourg (raphael.frank@uni.lu; sasan.jafarnejad@uni.lu).*

*Jiangbo Yu, Luis Miranda-Moreno, Jônatas Augusto Manzolli, Fuqiang Liu, Jiyao Wang, and Ali Eslami are with the Department of Civil Engineering, McGill University, Montreal, QC H3A 0C3, Canada (jiangbo.yu@mcgill.ca, luis.miranda-moreno@mcgill.ca; jonatas.manzolli@mcgill.ca; fuqiang.liu@mail.mcgill.ca; jiyao.wang@mcgill.ca; ali.eslami@mcgill.ca).*

## I. Introduction

Systems become agentic when they exhibit a degree of agency, understood as the capacity to form, interpret, and pursue self-generated or externally derived goals through context-sensitive reasoning, adaptive action, interaction, and reflective evaluation. The term *agency*, derived from the Latin *agere* ("to do" or "to act"), generally denotes the capacity to act intentionally toward desired outcomes. In philosophy, agency is commonly associated with acting on the basis of reasons, deliberation over alternative actions, and questions of responsibility and accountability. [1], [2]. In psychology, especially within the social cognitive tradition, agency is characterized through mechanisms such as forethought, self-regulation, self-efficacy, observational learning, and self-reflection, which together shape or govern goal-directed behavior [3], [4], [5]. Across these traditions, agency is understood as a purposive and context-dependent capacity to perceive conditions, reason about alternatives, make decisions, and pursue goals within evolving environments.

Agency is closely related to, yet distinct from, autonomy. Autonomy refers to the capacity to operate according to internally defined rules or learned policies independent of external control, whereas agency concerns how goals are formed, interpreted, and adaptively pursued in dynamic, often social contexts. Autonomous systems emphasize self-contained execution; agentic systems emphasize goal-oriented reasoning, interaction, and adaptation. Accordingly, a system may exhibit high autonomy with minimal agency, executing predefined routines, or substantial agency with limited autonomy, actively shaping goals and strategies while relying on external support for execution.

In recent years, agentic capabilities in engineered systems have undergone a notable improvement, driven by advances in Generative AI technologies, particularly Transformer-based large language models (LLMs), together with the enabling and supporting infrastructure, software architectures, configurations, and data ecosystems [6], [7], [8]. Generative AI primarily produces content in response to external prompts. In contrast, existing agentic systems embed generative models within a broader system architecture. They may maintain objectives over time, decompose tasks, interact with tools and environments, take actions, and adapt based on feedback. In this sense, Generative AI serves as an important enabling component, whereas agentic systems represent closed-loop, goal-directed processes that can decide, act, and learn in complex environments [6], [9], [10]. Furthermore, LLM-powered agentic systems are transitioning from purely digital settings to embodied and cyber-physical contexts [9], [11], where developments such as Application Programming Interfaces (APIs), Model Context Protocol (MCP), and cloud–edge computing enable real-time coordination across machines, organizations, and the built environment [12].

Since the initial preprint of this paper [13], we have also observed an emerging phenomenon in which industry actors are beginning to conceptualize—and in some cases experiment with—agentic AI capabilities in both manual and automated vehicles [14], [15], [16], [17], [18]. We refer to such an integration with vehicular systems as *agentic vehicles (AgVs)*. Fig. 1 illustrates this integration, together with the broader socio-technical interactions among humans, other machines, and the surrounding environment.

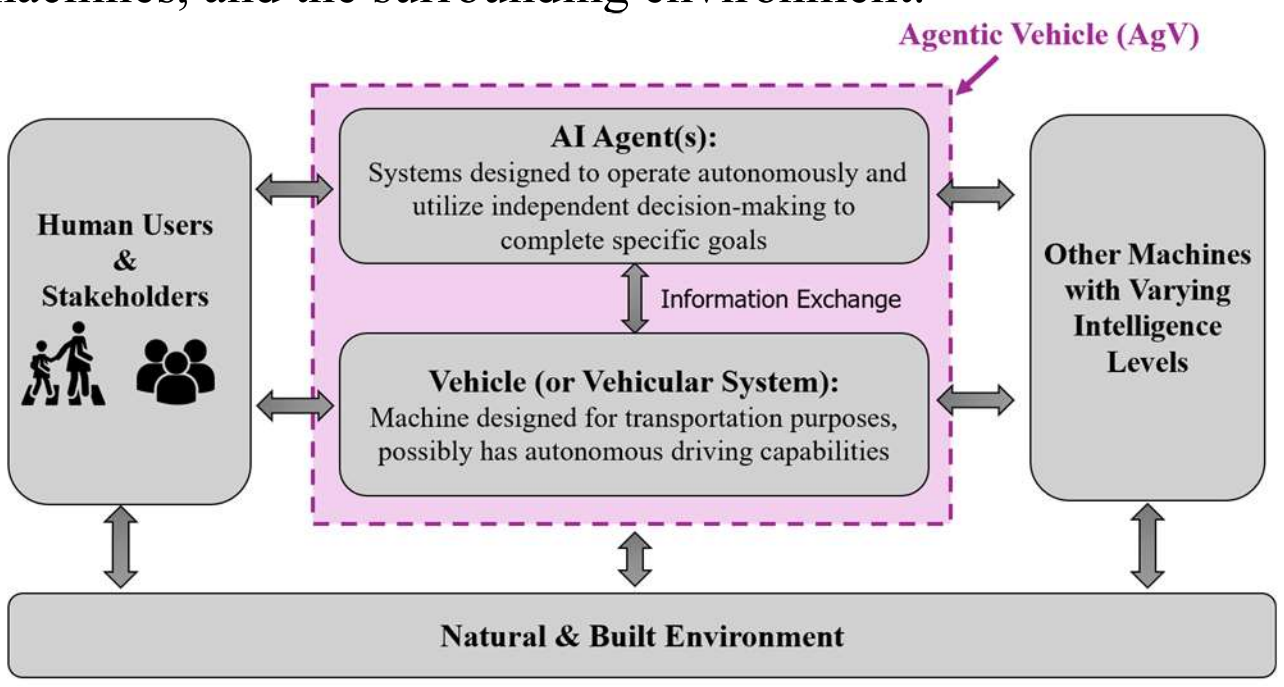

**Fig. 1.** An AgV emerges from the integration of one or more AI agents with a transportation machine, enabling more intelligent interactions with humans, other machines, and the environment.

This paper advances the conceptual and analytical foundations of next-generation intelligent vehicles. It introduces vehicle agency as a distinct dimension of vehicular intelligence, separate from the conventional focus on driving performance. We propose and formalize the concept of AgVs, develop illustrative use cases, establish the orthogonality between autonomy and agency, and outline pathways for their coordinated evolution. The proposed framework further elucidates system-level implications and policy considerations, supporting the secure, transparent, and human-centered integration of agentic AI, and its vehicular embodiment, into future transportation systems and the broader society.

## II. Agentic Vehicles

### *A. Definition*

We propose the following definition of an AgV:

> An **Agentic Vehicle (AgV)** is a vehicular system that exhibits a degree of *agency*— the capacity to interpret or formulate goals, reason across alternative courses of action under contextual uncertainty, engage and coordinate with human or machine actors, adapt and regulate its behavior through learning or reflective processes, and extend its functional and decision-making capabilities by leveraging external agents, tools, and resources.

This definition draws on both its foundational treatment in philosophy and psychology [1], [3] and its emerging usage in technology. In contemporary industry discourse, agentic AI is typically characterized by the ability of artificial agents to pursue complex goals through adaptive decision-making and a degree of operational independence [6], [7], [19]. Such accounts often position agency as a continuation of automation and autonomy along a developmental continuum, progressing from predefined task execution to systems capable of deliberation, adaptation, and goal formation. However, this interpretation underplays an important conceptual shift: beyond a certain point, the integration of intention management, context-sensitive reasoning, and socially situated interaction yields properties that are not merely incremental, but qualitatively distinct. At this point, it is more precise and analytically productive to recognize a new category rather than to extend existing ones.

Within this perspective, it is also important to distinguish the autonomy and agency of the AI agent from those of the vehicle it augments. While the AI agent's autonomy and agency pertain to its internal processes of goal interpretation, reasoning, and adaptive decision-making, the vehicle's autonomy and agency emerge at the system level through the coupling of these cognitive capabilities with physical actuation, sensing, and embedded operational constraints.

Regardless of whether emphasis is on its psychological and philosophical foundations or its technological instantiation, AgVs represent a distinct class of vehicular systems. They are not defined by their level of automation, but by the presence of mechanisms that support reasoning, deliberation, adaptation, and meaningful engagement with human actors as well as with natural and built environments. This formulation accommodates both fully automated and human-driven vehicles, thereby expanding the conceptual design of space beyond autonomy-centered paradigms.

### *B. Cognitive Mechanisms and Technological Foundations*

The conceptualization of AgVs can benefit from a three-layer framework that distinguishes externally observable capabilities from the components and infrastructures that enable them: **(i) agentic capabilities**—e.g., goal formation and interpretation, contextual and ethical reasoning, social engagement, self-reflection, external resource leveraging—describing what the system does at a functional level [9], [10], [20]; **(ii) cognitive functions**—e.g., sensing data processing, natural language processing, memory, learning algorithms—capturing the cognitive elements that realize these capabilities [8], [20]; and **(iii) foundational technologies and infrastructure**—e.g., electric power systems, sensors, vision-language-action (VLA) models, communication protocols, Vehicle-to-Everything (V2X) communication devices, computational cooling systems—providing the physical and digital substrate within which these components operate. This framework is not intended as a fixed architecture, but as an analytical scaffold for characterizing and designing agentic vehicular systems.

While many elements in the lower two layers overlap with conventional autonomous driving stacks [21], they do not necessarily enable agency. Agency arises from the coordinated orchestration of these components and infrastructures. An analogy can be drawn from assistive technologies: individuals with limited physical control may still exercise agency through mediated interfaces (e.g., eye gaze or gesture), provided that intentions can be formed and enacted. Likewise, agentic behavior in vehicles can exist under constrained execution capabilities, as long as intent can be represented and operationalized through available mechanisms.

### C. *Vehicle Agency Levels*

Agency should be best understood as a continuum rather than a binary attribute. Table I presents one possible illustrative classification framework. At Level 0, vehicles execute fixed, pre-programmed tasks without goal representation or contextual interpretation. At Level 1, vehicles become context-aware, adapting behavior to environmental conditions and supporting simple user-specified intentions. At Level 2, vehicles exhibit dialogic capabilities, interpreting high-level user goals and adjusting plans accordingly. At Level 3, vehicles demonstrate adaptive agency through multi-turn interaction, intent clarification, decision explanation, and dynamic plan revision. At Level 4, vehicles exhibit more advanced ethical, social, and reflective agency, including the ability to formulate and regulate goals, navigate trade-offs, coordinate with external systems, and justify decisions.

TABLE I. AN EXAMPLE VEHICLE AGENCY LEVEL DESIGNATION

| Level | Description | Example Case |
|---|---|---|
| **0:** Non-Agentic | Performs pre-programmed tasks based on fixed rules without understanding context or user intent. No human interaction beyond mechanical control. | Basic cruise control, automated emergency braking. |
| **1:** Context-Aware Responder | Adapts behavior based on environmental and operational context (e.g., traffic, weather); can interpret goals or intentions. Human interaction is limited to predefined inputs or overrides. | Traffic-aware rerouting, adaptive cruise control, lane keeping under environmental constraints. |
| **2:** Dialogic Agentic | Understands and responds to high-level human goals or preferences using simple natural language or interfaces. Capable of adjusting plans or priorities based on inferred user intent. | Responds to “take me to a quiet place,” adapts drop-off based on user time constraints or preferences. |
| **3:** Adaptive Agentic | Engages in multi-turn, multimodal dialogue with humans to clarify ambiguous goals, explain plans, or reason through novel scenarios. Demonstrates theory-of-mind–like behavior (e.g., inferring unspoken preferences). | Explains why a detour is needed, offers alternatives during disruptions, negotiates stops or shared rides. |
| **4:** Ethical, Social, Reflective Agent | Navigates complex social interactions and ethical dilemmas. Can form its own goals (with guardrails). Understands societal norms and emotional cues, collaborates with external systems (e.g., agents, web tools), and justifies decisions transparently. | Detects its own conditions, contacts auto stores to schedule repairs, informs city infrastructure teams about road damage, and updates future protocols through interactive or observational learning. |

Table II presents an illustrative mapping of how key capabilities emerge and intensify across levels of vehicle agency, highlighting the distinguishing features of increasingly agentic transportation machines. This framework enables the development of new evaluation metrics that extend beyond traditional engineering performance to include socio-technical alignment, ethical reasoning, and relational intelligence. In doing so, it provides a structured basis for guiding both the development of AgVs and their integration into broader socio-technical systems.

TABLE II. AN EXAMPLE SUMMARY AGV CAPABILITIES FOR EACH AGENTIC LEVEL

| Agentic Capability | AgV-0 | AgV-1 | AgV-2 | AgV-3 | AgV-4 |
|---|---|---|---|---|---|
| Context awareness | ✗ | ✓ | ✓ | ✓✓ | ✓✓✓ |
| Goal adjustment | ✗ | ✗ | ✓ | ✓✓ | ✓✓✓ |
| Natural language use | ✗ | ✓ | ✓✓ | ✓✓✓ | ✓✓✓ |
| Multimodal interaction | ✗ | ✗ | ✓ | ✓✓ | ✓✓✓ |
| Ethical reasoning | ✗ | ✗ | ✓ | ✓✓ | ✓✓✓ |
| External tool use | ✗ | ✗ | ✓ | ✓✓ | ✓✓✓ |

Note: ✗ indicates absence of the capability. ✓, ✓✓, and ✓✓✓ indicate increasing degrees of agentic capability.

## III. EARLY-STAGE IMPLEMENTATIONS AND PROSPECTS

Through reviewing recent relevant literature, we uncover three major application dimensions that utilize agentic capabilities for improving vehicle intelligence, including driving, social–contextual reasoning, and self-reflective management.

Agentic AI can enhance **semi- and fully-autonomous driving** [21], [22] by moving vehicle intelligence beyond predefined assistance functions toward context-aware, goal-directed, and interactive decision-making. Rather than merely detecting lanes, obstacles, nearby vehicles, and traffic conditions, agentic systems may reason about complex driving contexts, infer higher-level objectives, coordinate with drivers and other road users, adapt to changing conditions, and explain their actions. Recent work on multimodal foundation models and vision-language-action (VLA) models demonstrates this shift by supporting richer semantic understanding, language-conditioned reasoning, and closer alignment between driving decisions and high-level contextual objectives [23]. However, these capabilities are likely to augment rather than replace conventional autonomy stacks, at least in the near term. Their deployment also raises significant challenges related to verification, safety assurance, accountability, cybersecurity, human role, and the opacity of emergent decision-making behaviors.

In addition to driving, **social and contextual reasoning** allows the incorporation of human intent and evolving environmental situations [24], [25], [26], [27], [28], [29], [30]. Vehicles increasingly infer human states [30], [31], interpret high-level body and natural-language instructions, and adapt objectives in real time—expanding the tasks from solely driving into a continuously updated social cognitive process [32]. Such capabilities also introduce new uncertainties, as both human intent and environmental context must be inferred and validated under safety constraints [33], but it also enables behaviors such as adaptive routing, contingency handling, and cooperative coordination to emerge naturally from a unified reasoning loop. Communications are essential in most scenarios; however, when there is no response, AgVs can and should infer the situation and utilize external tools and resources to address the issue, as illustrated in Fig. 2.

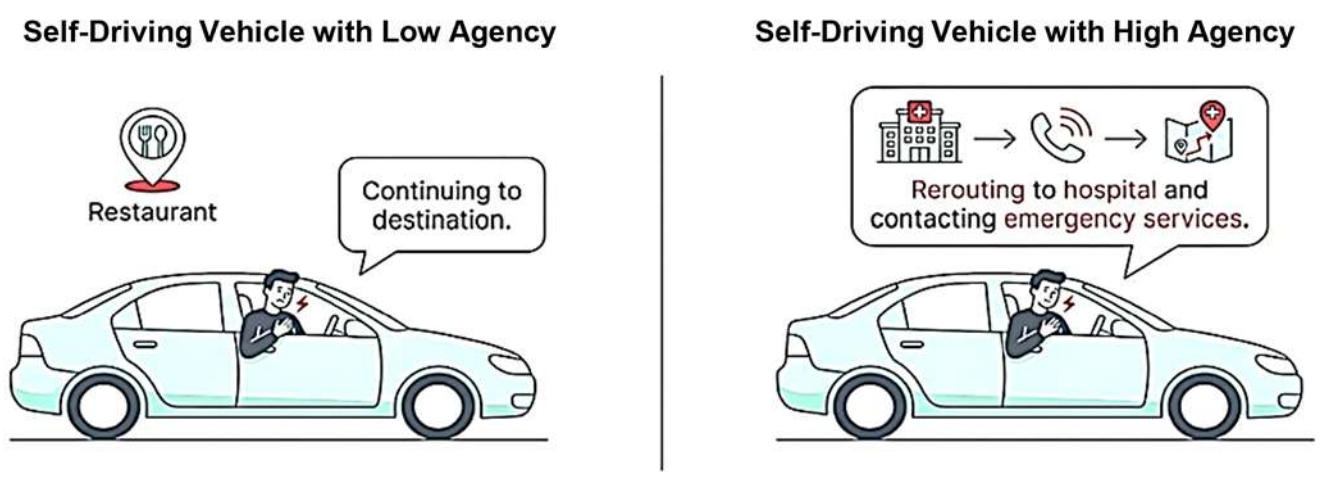

**Fig. 2.** Example scenario of a passenger having a heart attack in an AuV with low agency and one with high agency.

A third dimension directs decision-making inward through **self-assessment, regulation, and introspection**. AgVs may monitor, anticipate, and regulate their own internal states—such as battery status, engine health, sensor reliability, and system confidence (self-efficacy)—and use these to anticipate constraints and adapt behavior proactively. This includes predictive maintenance, energy-aware routing, and risk-sensitive fallback strategies based on uncertainty estimates [34], [35], [36]. Such capabilities transform diagnostics from static monitoring into dynamic, context-aware regulation, allowing the system to reason not only about the external world and human needs, but also about its own capabilities and limitations.

The early-stage attempts and their readily foreseeable extensions already suggest the emergence of a holistic vehicular intelligence ecosystem in which different agentic capabilities mutually reinforce one another, as illustrated in Fig. 3. While individual agentic capabilities may independently improve specific vehicular functions, their integration within a unified adaptive decision system may generate strong nonlinear and multiplicative effects across the vehicle lifecycle. Through such integration, vehicles may evolve from reactive transportation devices into continuously adaptive systems capable of self-monitoring, self-diagnostics, predictive maintenance, operational optimization, personalized interaction, cybersecurity monitoring and response, and context-aware behavioral adaptation under changing operational and environmental conditions. This shift suggests that future vehicles may increasingly function as persistent, self-managed, and lifecycle-aware intelligent systems rather than just isolated autonomous mobile units.

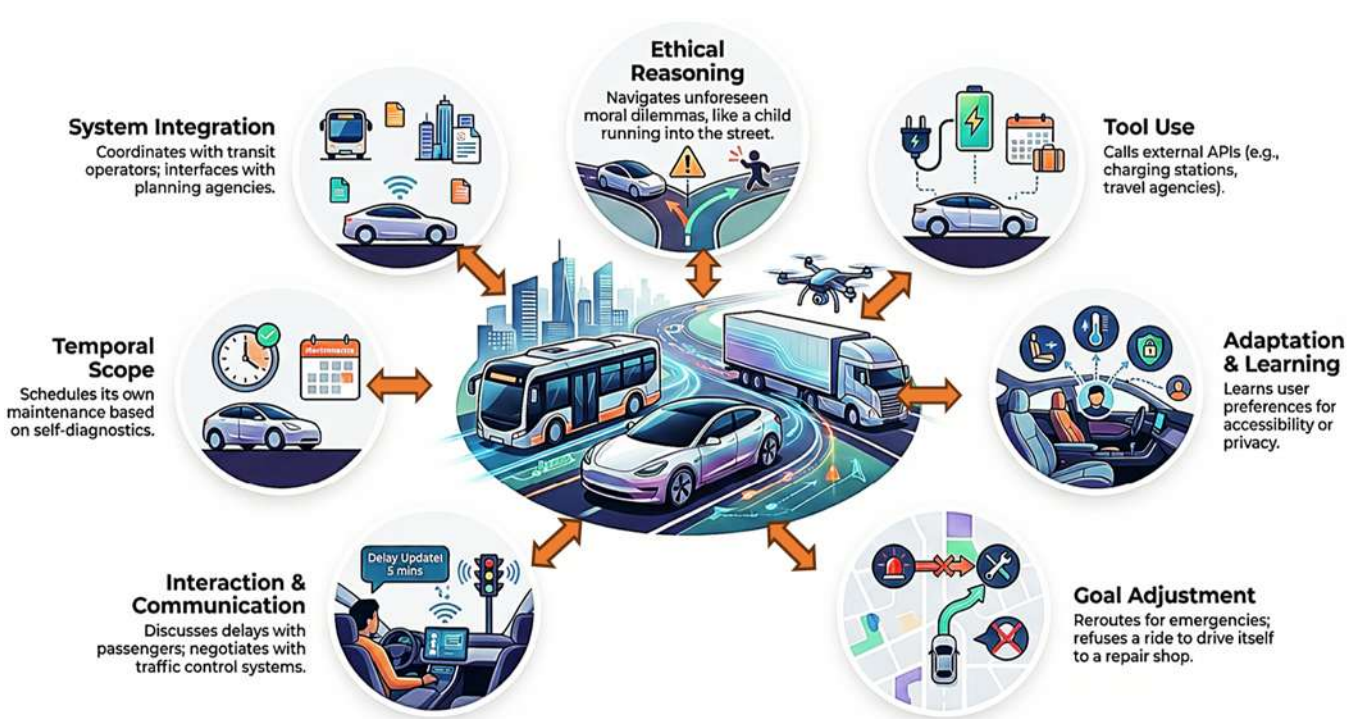

**Fig. 3.** Multimodal AgVs (e.g., cars, buses, trucks, shuttles, drones, and delivery robots) may integrate multiple agentic capabilities, including driving, intent management, social engagement, and self-assessment, to become more adaptive, self-managed, and lifecycle-aware intelligent vehicular systems.

## IV. Vehicle Agency and Autonomy Co-Development

Building on the distinction between agency and autonomy, intelligent vehicles should be characterized beyond a single axis of automation. Existing frameworks primarily place systems along a continuum of autonomy, capturing how tasks are executed but offering limited insight into how goals are interpreted, adapted, and coordinated.

### *A. Vehicle Agency–Autonomy Matrix (VAAM)*

To address this limitation, we introduce the VAAM, a two-dimensional framework defined by vehicle agency and vehicle autonomy, as illustrated in Fig. 4. To further clarify, the space is divided into nine principal autonomy–agency configurations, ranging from low-agency/low-autonomy systems to high-agency/high-autonomy systems. The framework emphasizes that autonomy and agency can be viewed as orthogonal dimensions: a vehicle may exhibit strong autonomous operational capability with limited agency, or advanced reasoning and interaction capabilities while still relying on humans for physical control.

Importantly, higher levels of autonomy and agency are not inherently superior. While they may enable greater adaptability, coordination, and operational capability, they may also introduce additional risks related to safety, controllability, transparency, accountability, cybersecurity, and human oversight. The VAAM therefore provides a structured basis for both classifying intelligent vehicles and evaluating the trade-offs, governance challenges, and societal implications associated with different autonomy–agency configurations.

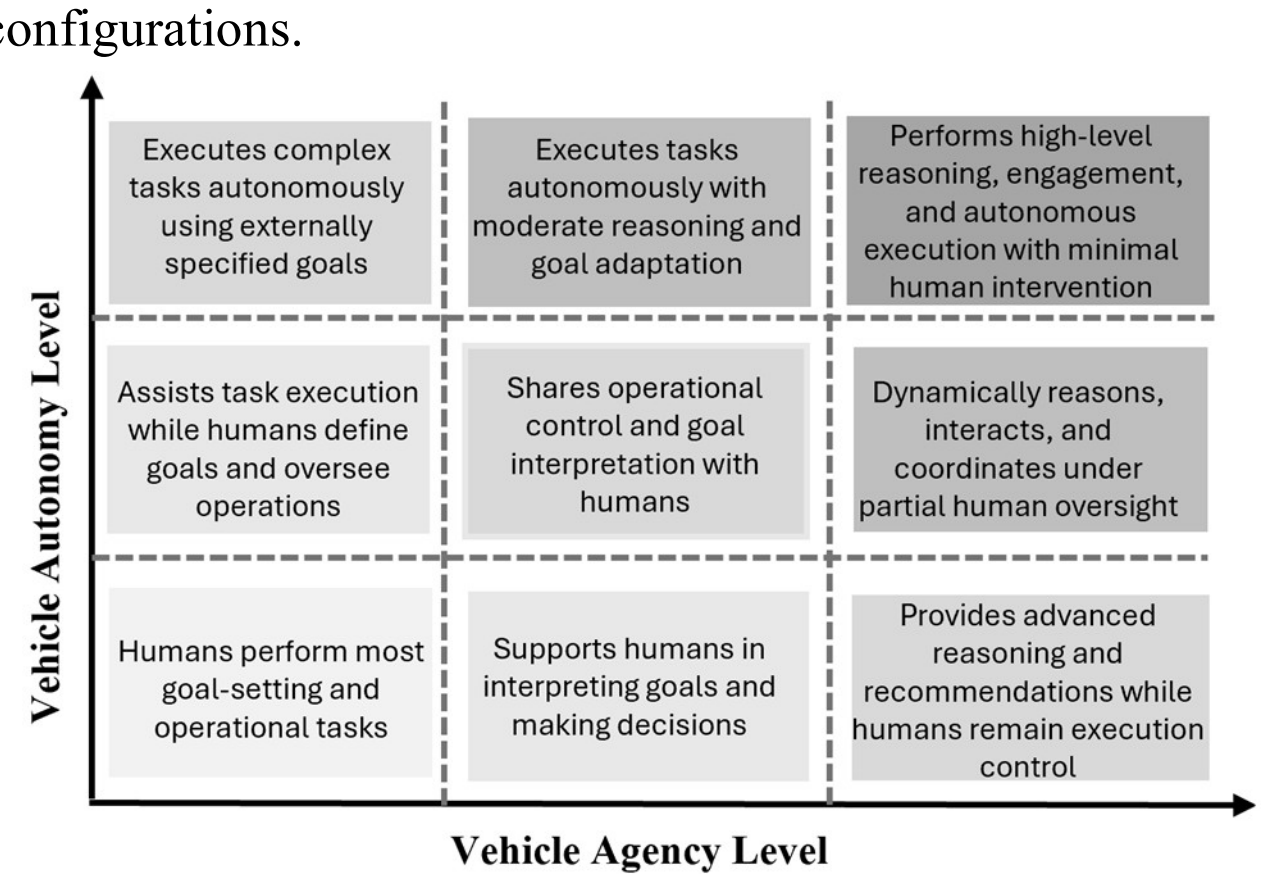

**Fig. 4.** Two-Dimensional Agency–Autonomy Framework for Analyzing, Evaluating, and Improving Vehicle Intelligence.

As an immediate application, the VAAM can help clarify how human driver roles evolve across different levels of vehicle intelligence. At low autonomy and agency, humans function as manual drivers without AI support. Increasing agency without substantial autonomy gives rise to informed or instructed driving, where AgVs provide guidance, recommendations, or alerts while humans retain direct control. At intermediate levels, roles shift toward lead, assisted, or supervised driving, with control and decision-making increasingly shared between humans and AgVs. At high levels of both autonomy and agency, humans transition into supervisors, collaborators, or strategic orchestrators, interacting through high-level goals, constraints, and coordination rather than continuous low-level control.

## B. Complementarities, Analogies, Use Cases, and Caveats

The most transformative vehicle intelligence may emerge from an appropriate integration of autonomy and agency, functioning not merely as automated machines but as cognitively capable and socially intelligible participants within mobility ecosystems. The agentic capabilities shape high-level objectives and strategies, while classical control executes commands and ensures operational feasibility, stability, and safety.

The complementarity of these two dimensions can be clarified through two disciplinary analogies. From a neurobiological standpoint, autonomy may align more with evolutionarily older brain structures responsible for motor control and routine execution, processes that are largely automatic in nature. In contrast, agency may correspond more to the prefrontal cortex, which supports reasoning, planning, and adaptation to novel contexts [37]. From a cognitive psychological perspective, autonomy may be analogous to System 1 processes (fast, automatic, and efficient), while agency parallels System 2 processes (deliberative, reflective, and goal-directed). Just as the interplay between subcortical systems and the prefrontal cortex, and the dynamic balance between System 1 and System 2, enable humans to combine efficiency with flexibility, the integration of autonomy and agency allows vehicles to transcend rote automation and engage in context-sensitive, goal-directed decision-making [38]. Importantly, treating one as inherently superior to the other overlooks the essence: It is their synergy, rather than the attempt to replace one with the other, that enables general vehicle intelligence.

An admittedly more speculative perspective further highlights the potential value of this synergy. Human drivers typically require only tens of hours of supervised and unsupervised practice to attain reasonable competence, albeit with limited scalability, whereas high-level AuVs demand millions of miles of driving data, corresponding to hundreds of thousands of hours of exposure, to approach comparable performance benchmarks [39]. This striking discrepancy suggests that autonomy alone, conceived as the large-scale execution of sensorimotor routines, is insufficient. What humans bring, and what current autonomous systems lack, may be agency: the capacity to generalize across contexts, reason about goals, and adapt strategies with limited exposure. That is, agency could be the "missing ingredient" that enhances and transforms autonomy, enabling learning efficiency and robust performance to reach a new level.

The opportunities and risks of combining autonomy and agency become evident in concrete cases where autonomously-driven AgVs (AAVs) can extend far beyond point-to-point driving to support complex, multi-stakeholder, and context-sensitive tasks. **In multi-step task orchestration**, an AAV might enable sequential school pickups, grocery retrievals, and synchronized passenger arrivals while negotiating with retailers, delivery platforms, or transit systems and adapting to changes expressed in natural language [40], [41]. These capabilities, however, raise concerns about equitable prioritization, exposure to manipulation, and the need for verifiable logs of how competing demands were weighed and resolved. **In proactive self-management**, an AAV may conduct continuous self-diagnostics, forecast its maintenance needs, and schedule charging or service appointments in coordination with fleet operators and grid conditions [42]. Yet, poorly calibrated reasoning could lead the vehicle to override user preferences, misinterpret operational constraints, or negotiate service schedules that inadvertently disadvantage other fleet assets. **In emergency scenarios**, an AAV may detect anomalous passenger states, contact emergency services, and reroute to appropriate facilities [43], [44], [45]. In specific cases, agentic reasoning might also cause the vehicle to prioritize saving the greatest number of human lives rather than the life of its human owner. Further, the vehicle might identify a pedestrian on the sidewalk showing signs of cardiac distress or detect a robbery in a nearby storefront and autonomously alert authorities, reposition itself as a source of safety or evidence, or coordinate with other vehicles to form access corridors for responders. These actions may introduce nontrivial risks and controversies, not to mention the potentially incorrect inferences that trigger false alarms, interfere with law enforcement operations, or expose sensitive data. Therefore, any autonomous intervention in public emergencies must be bounded by clear legal and ethical constraints. At the network scale, AAVs may act as critical components and players in **traffic estimation and management** [46]. Furthermore, AAVs can also act as privacy-preserving proxies for user mobility preferences, contributing aggregated insights to **long-term investment planning or policy-making processes** such as congestion pricing, curb-use regulation, or emergency-routing design [45], [47], [48], [49]. However, this role may also carry governance challenges around representational legitimacy, potential bias in aggregated data, and safeguards against unintended influence on public decision-making.

## C. Co-Development Pathways

The VAAM framework conceptualizes autonomy and agency as orthogonal dimensions, allowing multiple co-developmental trajectories rather than a single linear path. As illustrated in Fig. 5 (a–d), four broad strategies emerge (agency-first, autonomy-first, and two zig-zag co-evolutionary pathways), each shaping system architecture, human roles, and governance differently. Importantly, desirable outcomes need not correspond to full autonomy and full agency; intermediate configurations may be optimal depending on technical feasibility, regulatory context, and application domain.

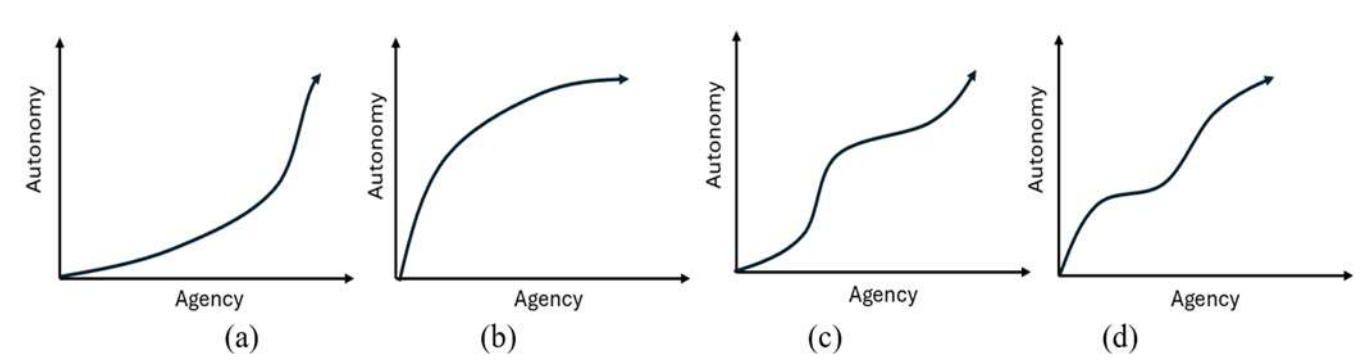

**Fig. 5.** Example co-developmental strategies

In the agency-first pathway (Fig. 5a), vehicles remain largely human-driven, while the agentic layer matures into a reasoning partner. Early systems emphasize contextual decision support, communication, and explainability—such as multimodal planning or infrastructure negotiation—placing human–vehicle interaction at the center. Accordingly, validation focuses on alignment, interpretability, and robustness of reasoning rather than control safety.

In contrast, the autonomy-first pathway (Fig. 5b) prioritizes perception, planning, and control, leading to highly automated systems with limited reasoning or communicative capability. These systems can operate safely in constrained environments but lack transparency and deliberative capacity. The key challenge then becomes retrofitting agentic functions, such as explanation, negotiation, and ethical reasoning, onto an already mature autonomy stack, with governance initially centered on safety and liability.

The co-evolutionary pathways (Fig. 5c–d) follow zig-zag progressions in which autonomy and agency advance in alternating and mutually enhancing steps. One variant introduces early agentic features before automation, gradually shifting from assistant to collaborator; the other begins with intermediate autonomy and subsequently integrates richer reasoning and interaction. Both require modular architectures, adaptive interfaces, and staged regulatory frameworks that accommodate incremental capability expansion.

Across all four example trajectories, the central implication is that development pathways jointly determine not only technical capabilities but also human roles and governance structures. Autonomy reduces direct human control, while agency reshapes interaction, accountability, and trust. Progress must therefore be evaluated beyond conventional performance metrics to include transparency, value alignment, and institutional legitimacy, with multiple trajectories likely coexisting across different markets, regulatory environments, and political contexts.

## V. System Implications

The emergence of AgVs may generate immediate system impacts that arise directly from their deployment in vehicles and mobility operations. By embedding cognitive, contextual, and interactive capabilities into vehicles, AgVs can influence vehicle control, user interaction, safety management, and real-time coordination, enabling more adaptive, personalized, and goal-responsive mobility services. However, these capabilities also introduce new risks related to operational safety, cybersecurity, explainability, human oversight, and value-laden decision-making [50]. In particular, agentic reasoning may create vulnerabilities associated with hallucination, misalignment, goal drift, adversarial manipulation, overreliance by human users, and uncertain human–machine interaction [51]. These immediate impacts highlight the need for staged deployment, continuous auditing, robustness testing, transparent communication, fail-safe mechanisms, and adaptive regulatory oversight.

AgVs may also generate wider indirect system impacts that emerge through broader behavioral, economic, environmental, spatial, and institutional responses. Optimized routing, cooperative driving, and adaptive fleet operations could improve network efficiency, safety, energy use, emergency response, and accessibility. However, increased convenience, lower perceived travel burden, and more flexible mobility services may also induce household and business relocation, additional vehicle kilometers/miles traveled (VKT/VMT), and the redistribution of risks across communities and infrastructure systems. AgVs may further disrupt labor markets by displacing some driving and logistics occupations while creating new roles in AI oversight, fleet orchestration, system auditing, human–machine coordination, and regulatory compliance. Their long-term societal value will therefore depend on interoperable standards, privacy protection, participatory governance, institutional accountability, and alignment with public values [9], [52], [53], [54], [55].

AgVs may also play a key role across mobility system functions and lifecycle phases, as illustrated in Fig. 6. Beyond serving as adaptive and self-managed vehicular systems, AgVs may increasingly function as distributed nodes for sensing, communication, coordination, and decision-making within broader socioeconomic systems. This role is consistent with emerging applications and visions of agentic AI in long-term public investment, strategic transportation planning, transit and ridesharing fleet procurement, resident-oriented mobility support, sensor deployment and data collection, and vehicle/fleet finance and pricing. [9], [47], [48], [56], [57], [58], [59], [60], [61], [62]. Through these integrations, AgVs may support transportation systems across multiple lifecycle phases, from long-term planning, system analysis, investment evaluation, and infrastructure design to real-time monitoring, adaptive operational control, maintenance management, emergency response, and user-facing travel assistance. Collectively, these developments suggest a transition from isolated or locally connected intelligent mobile units toward interconnected, continuously adaptive transportation ecosystems, with significant implications for system architecture, interoperability, data governance, cybersecurity, accountability, institutional coordination, and public oversight.

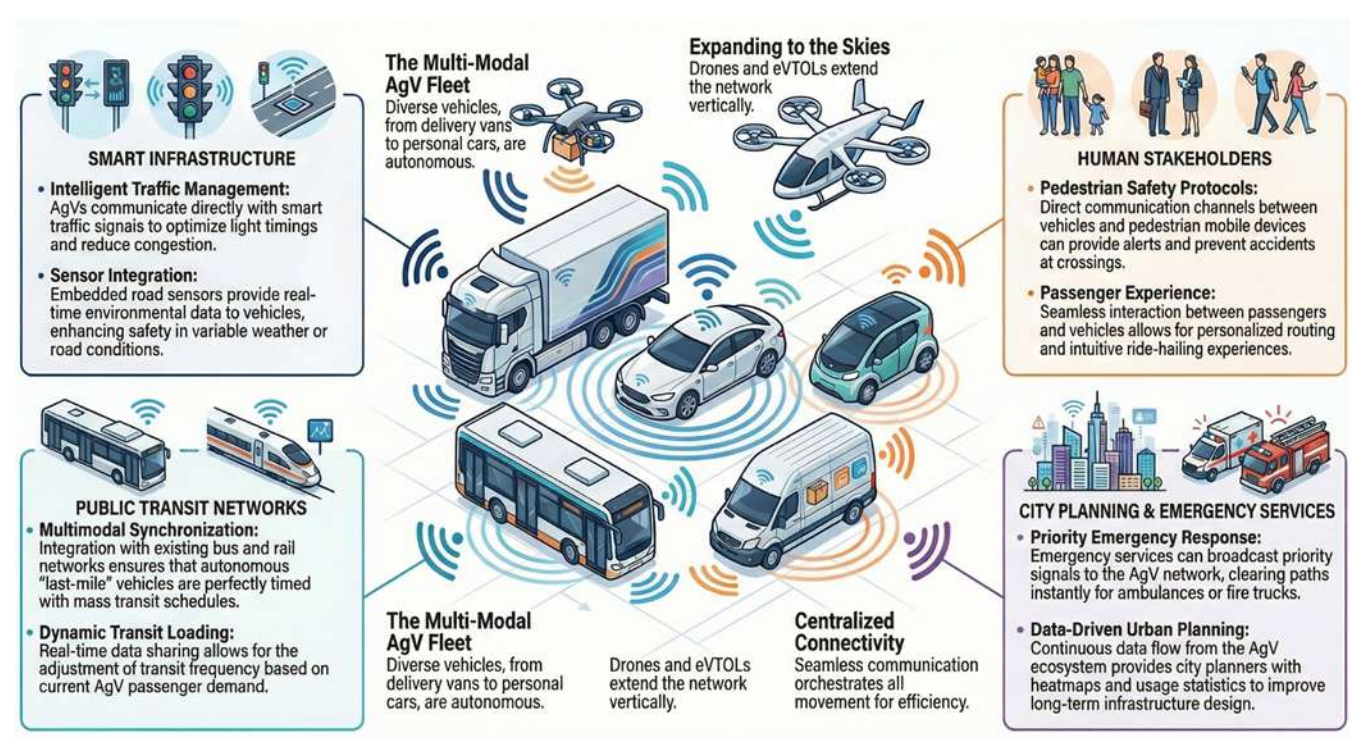

**Fig, 6.** Illustration of the agentic mobility ecosystem, in which AgVs (e.g., agentic cars, trucks, drones, buses, vans) can interact with pedestrians, the built environment, institutions, and other agentic machines, extending beyond an intelligent cockpit.

From a systems perspective, vehicles constitute a particularly suitable embodied edge platform for deploying agentic intelligence in both routine and emergency public-use scenarios. Compared with devices such as smartphones, many vehicles offer greater computational capacity, energy availability, physical space, sensor integration, and operational robustness, enabling partial or full local execution of AI agents without continuous reliance on cloud infrastructure. This capability can strengthen resilience under limited or disrupted connectivity while positioning vehicles as mobile hosts of agentic intelligence, rather than passive endpoints of remote computation.

## VI. CONCLUSIONS

This paper reframes intelligent vehicles by introducing agency as a dimension distinct from, yet complementary to, autonomy. While autonomy concerns the reliable execution of tasks, agency concerns the capacity to interpret goals, reason under uncertainty, adapt to context, and interact meaningfully within dynamic socio-technical environments. This distinction provides a clearer foundation for understanding and guiding the emerging capabilities of next-generation mobility systems.

AgVs thus represent a shift from passive transportation tools toward adaptive participants in mobility systems. Their behavior is shaped not only by control and navigation performance, but also by how they formulate or interpret objectives, interact with users and infrastructure, and respond to evolving operational and social conditions. Different autonomy–agency configurations may therefore produce distinct operational roles, risks, and system-level outcomes.

Realizing the potential of AgVs will require advances not only in engineering, but also in system architecture, modeling, governance, and public oversight. Future mobility systems must ensure transparency, alignment, cybersecurity, and accountability within increasingly distributed and interactive decision processes. Ultimately, the challenge is not only to develop more capable vehicles, but to integrate them responsibly into mobility ecosystems. The long-term success of AgVs will depend on aligning technological development with societal values to create mobility systems that are efficient, adaptive, trustworthy, and human-centered.